\documentclass[sigconf]{acmart}
\usepackage{xcolor, soul}
\usepackage{subfigure}
\sethlcolor{green}
\AtBeginDocument{%
  \providecommand\BibTeX{{%
    \normalfont B\kern-0.5em{\scshape i\kern-0.25em b}\kern-0.8em\TeX}}}

\begin{document}

\begin{figure}
 This paper has been accepted for publication in the 2023 Genetic and
Evolutionary Computation Conference.\\
To cite this work, please use the following citation:
Maryam Kebari, Annie S. Wu, and H. David Mathias (2023).  PID inspired
modifications in response threshold models in swarm intelligent
systems.  In the {\em Proceedings of the Genetic and Evolutionary
Computation Conference}.
\end{figure}
%%
%% The "title" command has an optional parameter,
%% allowing the author to define a "short title" to be used in page headers.
\title{PID-inspired modifications in response threshold models in swarm intelligent systems }

%%
%% The "author" command and its associated commands are used to define
%% the authors and their affiliations.
%% Of note is the shared affiliation of the first two authors, and the
%% "authornote" and "authornotemark" commands
%% used to denote shared contribution to the research.

\author{Maryam Kebari}
\affiliation{%
 \institution{University of Central Florida}
 \city{Orlando}
 \state{Florida}
 \country{USA}}
  \email{m.kebari@knights.ucf.edu}

\author{Annie S. Wu}
\affiliation{%
 \institution{University of Central Florida}
 \city{Orlando}
 \state{Florida}
 \country{USA}}
\email{aswu@cs.ucf.edu}

\author{H. David Mathias}
\affiliation{%
 \institution{University of Wisconsin - La Crosse}
 \city{La Crosse}
 \state{Wisconsin}
 \country{USA}}
\email{dmathias@uwlax.edu}

%%
%% By default, the full list of authors will be used in the page
%% headers. Often, this list is too long, and will overlap
%% other information printed in the page headers. This command allows
%% the author to define a more concise list
%% of authors' names for this purpose.

%%
%% The abstract is a short summary of the work to be presented in the
%% article.
\begin{abstract}
In this study, we investigate the effectiveness of using the PID (Proportional - Integral - Derivative)
control loop factors for modifying response thresholds in a decentralized, non-communicating, threshold-based swarm.
Each agent in our swarm has a set of four thresholds, each corresponding to a task the agent is capable of performing. The agent
will act on a particular task if the stimulus is higher than its corresponding threshold. The ability to modify their thresholds allows
the agents to specialize dynamically in response to task demands.
Current approaches to dynamic thresholds typically use a learning and forgetting process to adjust thresholds. These methods are able to effectively specialize once, but can have difficulty re-specializing if the task
demands change. Our approach, inspired by the PID control loop,
alters the threshold values based on the current task demand value,
the change in task demand, and the cumulative sum of previous
task demands. We show that our PID-inspired method is scalable
and outperforms fixed and current learning and forgetting response
thresholds with non-changing, constant, and abrupt changes in
task demand. This superior performance is due to the ability of our method to re-specialize repeatedly in response to changing task demands.
\end{abstract}

\copyrightyear{2023} 
\acmYear{2023} 
\setcopyright{acmlicensed}\acmConference[GECCO '23]{Genetic and Evolutionary Computation Conference}{July 15--19, 2023}{Lisbon, Portugal}
\acmBooktitle{Genetic and Evolutionary Computation Conference (GECCO '23), July 15--19, 2023, Lisbon, Portugal}
\acmPrice{15.00}
\acmDOI{10.1145/3583131.3590442}
\acmISBN{979-8-4007-0119-1/23/07}

%%
%% Keywords. The author(s) should pick words that accurately describe
%% the work being presented. Separate the keywords with commas.

%% A "teaser" image appears between the author and affiliation
%% information and the body of the document, and typically spans the
%% page.

%%
%% This command processes the author and affiliation and title
%% information and builds the first part of the formatted document.

\maketitle

\section{Introduction}

\par In this study, we use a PID-inspired method for manipulating agent response thresholds in a threshold-based swarm. The swarm is fixed-size, decentralized, and without communication. At any given time, an agent can perform at most one of the four possible tasks that may arise in the test problem. Current dynamic threshold methods can specialize once but can have a hard time respecializing. Our method helps the swarm to adjust to varying task demands more effectively by enhancing swarm's ability to re-specialize.
\par Decentralized swarms are robust and scalable, but not having a leader makes it challenging for the swarm to achieve a collective goal. The response threshold model is a simple method for addressing this problem. Natural swarms such as social insects have inspired this method \cite{bonabeau1996quantitative,bonabeau1998fixed,theraulaz1998response}. The simple yet powerful idea behind this model is that each agent has a set of thresholds, one for each task present. If the level of a task stimulus or task demand is below its corresponding threshold, the behavior is less likely to happen. The behavior is more likely to occur if the stimulus is above its corresponding threshold \cite{bonabeau1996quantitative,theraulaz1998response}. Hence, the thresholds we assign to each agent determine the system's collective behavior. We are looking for a suitable assignment of the threshold values to the agents for the current task demands. Any system needs to adapt to changes and new task demand requirements to survive and thrive. In response-threshold-based systems, this adaptation relies on each agent's ability to adjust its threshold as the experiment progresses.  If a specific task is currently in demand that was not before, the swarm needs more agents with low thresholds for that specific task in order to be able to respond effectively. 
Current methods for threshold adaptation can effectively specialize to an initial set of task demands, but often have difficulty respecializing if task demands change.
\par We aim to find a method that allows  the agents to specialize \emph{and} re-specialize effectively. A control loop can be helpful when addressing this problem. Control loops try to achieve the desired output by continuously adjusting the system based on feedback. \emph{PID control} (Proportional - Integral - Derivative) is a popular and straightforward control loop \cite{borase2021review}. It is widely used in various applications, such as particle biomedical applications \cite{slate1982automatic,isaka1993control}, process control such as temperature control \cite{yamamoto2004design}and, robotics \cite{zhang1990minimum}.

 \par A PID controller continuously compares the desired behavior of the system with its actual behavior, then utilizes that error rate to alter the system's behavior to get it closer to the desired behavior. This alteration depends on three components: The proportional component adjusts the control signal based on the current error rate. The integral component takes into account the accumulated error rate over time, which helps eliminate any steady-state error. The derivative component considers the change in error rate. This component can help the system's responsiveness in the face of sudden changes. By applying the PID controller to our system, we alter the thresholds of the agents based on these three components. 

\par We test our approach on five types of task demand setups with three swarm sizes. We demonstrate that our PID-inspired method is scalable and improves swarm performance compared to the current dynamic and fixed threshold methods.

\par

\section{Background and related work}

\subsection{Response threshold models}
\par Earlier studies show that fixed response threshold models \cite{bonabeau1996quantitative,kanakia2016modeling,krieger2000call,wu2018inter} work relatively well in both static and changing environments \cite{wu2020dynamic}. We contend that such fixed models do not exploit the full potential of those systems, and allowing thresholds to change over time in response to environmental or other factors can result in a more efficient system. Such changes might be non-reactive or reactive. In non-reactive methods, thresholds are a function of time. They do not depend on local or global factors of the experiment. An example method inspired by social insects \cite{robinson1987regulation} is age-polyethism, where agents' thresholds for tasks or their range change as they age \cite{vesterstrom2002division,merkle2004dynamic}. Reactive methods, on the other hand, alter thresholds when the agent reacts to something happening during the experiment. They could be based on local or global factors. Global factors refer to factors such as task demand, the number of agents acting on a particular task. Information about these factors might not be accessible to the agents, so sometimes, they need to estimate the global values based on their communication with other agents \cite{lee2019adaptive}. An agent's experience with a particular task \cite{theraulaz1998response,campos2000dynamic} in an example of a local factor.
One of the important instances of local factors is the use of learning/forgetting rates \cite{campos2000dynamic,cicirello2003distributed,price2004evaluation,theraulaz1998response}. The agent's threshold decreases by a constant value when the agent performs a task and gains experience and increases when the agent forgets a task by not performing it for a time step. This creates a positive feedback loop which can cause agents to become stuck at very high or low thresholds. Therefore, these approaches can adapt in the early stages but may become less effective as the experiment progresses \cite{kazakova2018specialization,kazakova2020respecializing}.
One example of this problem occurs when all tasks are not initially introduced. Agents become specialized on the initial tasks but have difficulty respecializing when new tasks are introduced.
We introduce a new method that lets the agent specialize and re-specialize without getting stuck with very low or very high thresholds.

\subsection{The PID controller}

\par \emph{Closed control loops} are mechanisms in which the system's error rate, defined as the difference between the setpoint ($SP$) and the process variable ($PV$), is continuously calculated. The controller then applies a correction based on that error rate to improve performance in each time step. Some examples of closed control loops are the PID controller \cite{xiang2019simple,slate1982automatic,isaka1993control,yamamoto2004design,zhang1990minimum}, fuzzy logic controller \cite{berenji1992fuzzy,pappis1977fuzzy}, model predictive controller \cite{rawlings2000tutorial,razzaghpour2021impact}, and adaptive controller \cite{seborg1986adaptive,anderson2008challenges,shahram2023energy}.

\par
\begin{figure}[h]
  \centering
  \includegraphics[width=\linewidth]{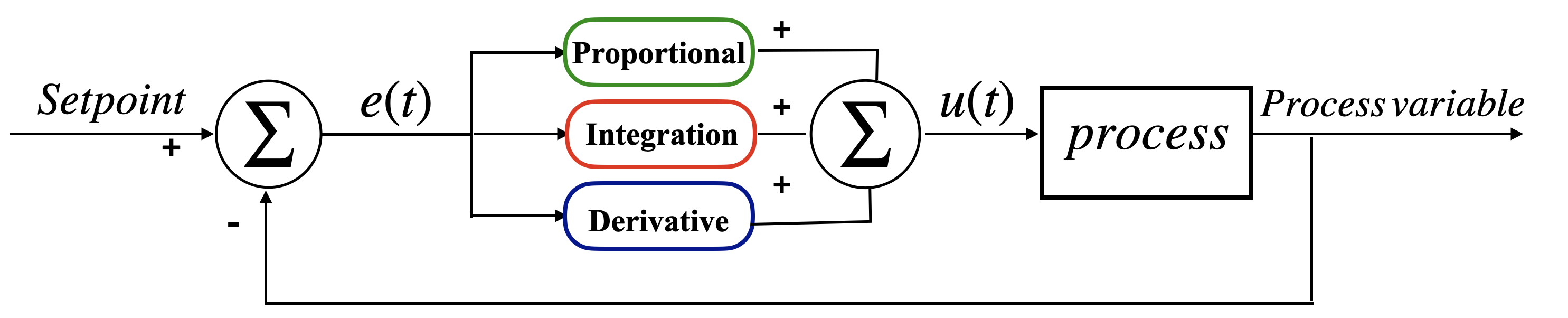}
  \caption{Block diagram of a standard PID controller}
  \label{fig:1}
  \Description{Block diagram of a standard PID controller}
\end{figure}

 The PID controller offers a simple solution to many control problems \cite{borase2021review}. The simplicity of the PID controller makes it a suitable choice for large-scale swarms, because, as the swarm size increases, high computational power for each agent might not be feasible and can be costly. Therefore, a simple controller that does not need high computational power is preferred.
 A block diagram of the PID controller is shown in Figure~\ref{fig:1}.

Figure~\ref{fig:1} shows that the difference between the setpoint and the process variable, which is the error rate, is fed into the controller.
The controller has three components, proportional, integral, and derivative. These three factors are summed up by a summing junction and then are processed by the plant, which is the system we are trying to control. Then the process variable is recalculated after the correction, and the loop continues.
The error rate, $e(t)$, given setpoint, $SP$, and process variable, $PV$,  can be written as :
\begin{equation}
  e(t)=SP(t)-PV(t)
\end{equation}
The output of the PID control function, $u(t)$, given error rate, $e(t)$, gain of P, $K_P$, gain of I, $K_I$, and gain of D, $K_D$,  is calculated by:
\begin{equation}
  u(t)=K_P.e(t)+K_I. \int_{0}^{t}e(\alpha )d\alpha +K_D\frac{de(t)}{dt}
\end{equation}
Gains $K_P$, $K_I$, and $K_D$ change based on the problem settings and are determined by tuning methods. Gains can be zero, which leads to the elimination of some terms.

\par The proportional term correlates the controller's response with the system's current error rate. Although this makes our system responsive, using the proportional term alone (P controller) can lead to a long settling time and a high steady-state error.

\par The integral term operates by looking at accumulated past errors.
Using the integral term alone (I controller) can lead to a slow response time; however, the I controller has less steady-state error than the P controller. Another problem with the integration factor is the windup problem. A windup occurs when the accumulated error from the integration factor keeps pushing the system even though the current error rate is low. Hence, the windup worsens the error and works against the benefit of the system; however, this issue can be addressed by anti-windup methods.
By combining the P and I controllers (PI controller), we improve the responsiveness compared to the I controller and reduce the steady state compared to the P controller.

\par The derivative term functions by considering the current trend of the system's error rate. The derivative term is not typically used alone because it makes the system highly unstable. It is combined with the proportional term (PD controller) or both proportional and integral terms (PID controller). The derivative term is used to increase the system's responsiveness. The PD and PID controllers have higher responsiveness than P and PI controllers, which can be helpful in systems with abrupt changes, although it amplifies noise and cannot eliminate steady-state error.

\par To gain more intuition, consider an example: a drone wants to hover at an altitude $L$ from the earth. If we use the proportional term alone, once the drone reaches altitude $L$, it stops pushing and falls. There is a certain point where the push of the drone is equal to the drone's weight, which is the steady-state error. The integral term can eliminate this. Steady-state error accumulates over time and causes the drone to push more: As long as there is an error, it accumulates and forces the drone to push; however, the accumulated error may cause the drone to pass the desired point. We need a term to determine whether we are closing in on our goal, which is the derivative term. 

\par Given the limitations described above, a controller that can manage most circumstances should have all three terms. 
\section{Experimental methods}

\subsection{Testbed setup}
\par

The testbed we use is a task allocation problem visualized as a collective tracking problem, frequently used in response threshold studies \cite{wu2021collective,wu2017effects,wu2018inter}.
\par
Our testbed is inspired by the thermoregulation of bees. A colony
of bees can collectively control the temperature of a hive over time without a centralized controller. The temperature of a
hive (a one-dimensional vector) rises when bees shiver and declines when they flap their wings \cite{weidenmuller2004control}; each bee has a choice between the two tasks (flapping or shivering), which are in opposition. We extend this to two dimensions and frame it as a collective control 2D tracking problem. This extends our model so it can model a more complex system and also makes visualization easier \cite{wu2021collective}.

We consider a moving target and a tracker in a 2D vector
space. In each time step, the tracker should follow the target as closely as possible. The performance of the system is measured as the average distance between the
target and the tracker over all time steps.
 The \emph{error rate} in our system is defined as the distance between the target and the tracker.
The target moves in a prescribed path in the 2D space for a fixed distance in each time step.
The tracker is controlled collectively by the swarm of agents, meaning each agent can choose one of the following; it can either push north,
push south, push west, push east, or remain idle. The actions of all agents are aggregated to produce the target movement in each time step. The swarm of agents is decentralized and agents are non-communicating.
Agents make the decision of what task to choose based on their thresholds and the current corresponding task demand. Task demand is defined as the difference between the target and the tracker in each direction at each time step. Each agent $i$ has a set of four thresholds $T_{i, north}$, $T_{i,south}$, $T_{i,east}$, $T_{i,west}$. If the task demand  for a certain direction exceeds the agent’s threshold for that direction, the agent will perform that task unless there exists more than one task with thresholds lower than the stimulus. In the latter case, the agent will choose one of them randomly. We run our experiments for 500 time steps to give our methods time to settle.

Although the visualization of our problem looks like a tracking problem and we realize there are very efficient algorithms that solve tracking problems. The goal of our study is not to find a more efficient method for the tracking problem, but rather is to find a method to change thresholds in a decentralized task allocation problem so that the system performs effectively in changing and non-changing environments. For this purpose, we need a testbed in which task demand changes could be made easily and visualized clearly. In this testbed, we can drastically change the task demand at each time step by changing the target path defined. Changes can happen gradually or suddenly, in multiple or just one direction, or everything can remain constant. This makes this testbed an excellent fit for our purpose. For example, in a zigzag path, demand to the east is constant, and demands to the north and south alternate abruptly. On the other hand, we have the s-curve path, a periodic path with constant east demand and continuously changing demand between north and south. Using these paths, we can test if our method is effective in both continuously and abruptly changing environments. We perform experiments on five paths.

\begin{itemize}
\item {\verb|west|}: In each timestep, there is only a constant demand for the west. This path is chosen to show how our system reacts to non-changing task demands.
\item {\verb|random|}: In each time step, a random angle from the Gaussian distribution N (0.0, 1.0) gets added to the current target direction. In almost all instances, random paths will involve demands in all directions.

\item{\verb|sharp|}: Involves travel in a straight line with a probability of turning in a random direction in each time step
\item{\verb|s-curve|}: A periodic path that has a constant demand for the east task, the demand for north and south continuously change, similar to a sine path. Requires all tasks, but the demand for the west direction is very small. 
\item{\verb|zigzag|}: A periodic path with constant east demand. the demand for north and south are constant for a period of time and then abruptly change, creating a zigzag pattern 
\end{itemize}
 We perform each experiment with 500 time steps to give time to the PID to settle and converge. 
\subsection{PID-inspired method}
\subsubsection{Model setup}
\par In our method, inspired by the PID control loop, thresholds change based on the swarm's error rate, derivative, and integral.

\par In our test bed, the error rate $e_{i,t}$ in each direction,  $i\in\lbrace N, E, S, W\rbrace$ at time step $t$, given target and tracker locations, $Target_{x,t}$,$Target_{y,t}$ and $Tracker_{x,t}$, $Tracker_{y,t}$  is defined as:
\begin{equation}
    e_{N,t} = Target_{y,t} - Tracker_{y,t} \quad \text{if $Target_{y,t} > Tracker_{y,t}$ }\\
    \end{equation}
    \begin{equation}
    e_{S,t} = Tracker_{y,t} - Target_{y,t} \quad \text{ if $Target_{y,t} < Tracker_{y,t}$}\\
    \end{equation}
    \begin{equation}
    e_{E,t} = Target_{x,t} - Tracker_{x,t} \quad \text{if $Target_{x,t} > Tracker_{x,t}$} \\
    \end{equation}
    \begin{equation}
    e_{W,t} = Tracker_{x,t} - Target_{x,t} \quad \text{ if $Target_{x,t} < Tracker_{x,t}$} 
    \end{equation}
and otherwise they are zero.
\par In each time step, $t$, given the proportional factor, $F_P$, integral factor, $F_I$, derivative factor, $F_D$, gain of P, $K_P$, gain of I, $K_I$ and gain of D, $K_D$, agent $i$'s threshold for task $n$, $T_{t,i,n}$, can be calculated by:
\begin{equation}
    T_{t,i,n} =T_{t-1,i,n}-K_P\times F_P-K_I\times F_I-K_D\times F_D
\end{equation}

We define the proportional factor, $F_{P,i,t}$ for $i\in\lbrace N, E, S, W\rbrace$, at time step $t$, to be:
\begin{equation}
F_{P,N,t}=e_{N,t} - e_{S,t}  
\end{equation}
\begin{equation}
F_{P,E,t}=e_{E,t} - e_{W,t} 
\end{equation}
\begin{equation}
F_{P,S,t}=e_{S,t} - e_{N,t}  
\end{equation}
\begin{equation}
F_{P,W,t}=e_{W,t} - e_{E,t}
\end{equation}
which is the error rate in each direction. Based on the error rate, $e_{i,t}$ definition, in each time step, either the north or south and either east or west direction error rates must be equal to zero.

We define the integral factor, $F_{I,i}$ given $i\in\lbrace N, E, S, W\rbrace$  , at time step $t$, to be:
\begin{equation}
F_{I,i,t}= \sum_{s=0}^{t-1} F_{P,i,s}
\end{equation}
which is the integral component written in discrete form by a summation because of the discrete time steps of our problem.

We calculate the derivative factor, $F_{D,i,t}$, for $i\in\lbrace N, E, S, W\rbrace$ using :
\begin{equation}
F_{D,i,t}=e_{i,t}-e_{i,t-1}
\end{equation}
$F_{D,i,t}$ is the difference between the two recent error rates in each direction.
\par
\begin{figure*}[t]
  \centering
  \includegraphics[width=6in]{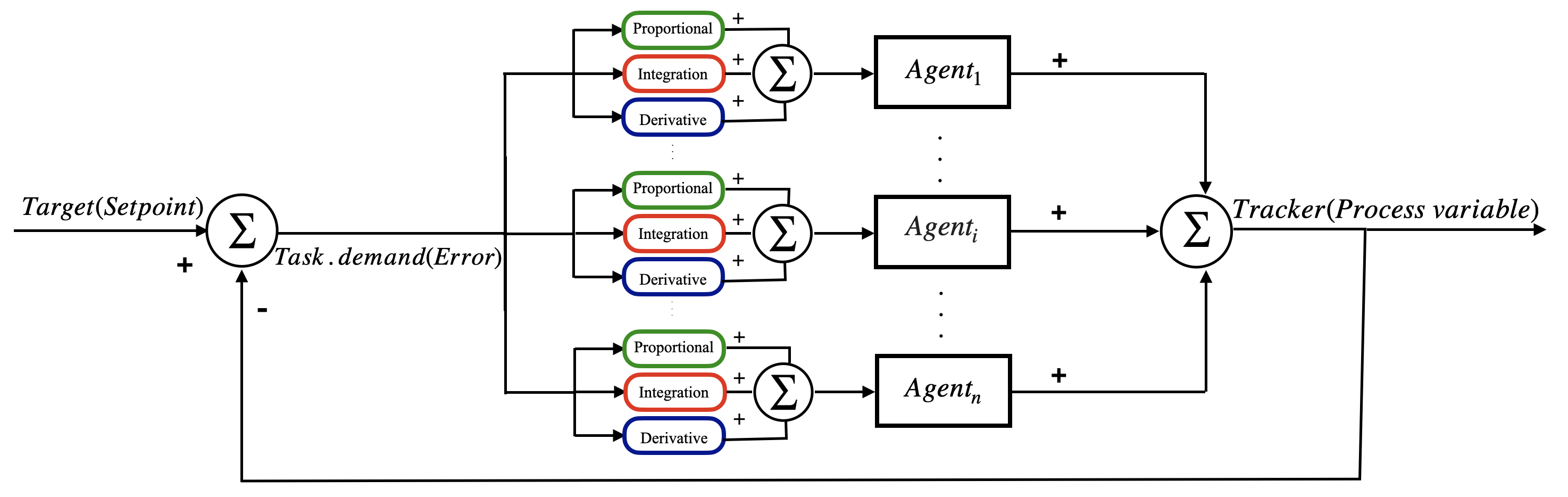}
  \caption{Block diagram of our PID-inspired method}
  \label{fig:2}
\end{figure*}

 Figure~\ref{fig:2} shows the block diagram of our method. Each agent $i$ has a separate controller. 
The difference between the target and tracker is calculated and fed to all controllers. After modifying the thresholds based on our method, the agents decide whether to act on that specific task. Their decision could be 1 or 0. One means the agent performs the task and zero means the agent refuses. The sum of the agent's decisions updates the tracker location, and the loop continues.

\subsubsection{PID tuning}
Determining the gains of a PID controller's proportional, integral, and derivative factors plays a vital role in how well the system performs \cite{borase2021review}.
Multiple methods are used for tuning, including the \emph{Ziegler-Nichols method} \cite{ziegler1942optimum}. The Ziegler-Nichols method is one of the simplest methods for tuning the PID controller. First, we set $K_D$ and $K_I$ to zero, then increase $K_P$ until our output has consistent oscillation, note this gain as $K_U$ and its oscillation period as $P_U$. Then based on the  Ziegler-Nichols's method, $K_p$, $K_I$ and $K_D$ can be written as \cite{ziegler1942optimum}:
\begin{equation}
K_P=0.6 K_U \quad
K_I=1.2K_U/P_U \quad
K_D=0.075s K_U\times P_U
\end{equation}
Since we might not always have prior knowledge about the path and task demands we are going to encounter, we choose to tune our parameters based on a basic straight path, and use the same parameters for other paths as well. 

\subsubsection{Integral windup}

The integral component and the accumulated error rate can reduce the steady state error, but this also has a downside. If the error, in the beginning, is too large, even though the current error rate is low, the error rate windups from the beginning of the experiment can make the integral component too large and make the performance worse. Accumulated errors from the previous time steps can cause the swarm to put in too much effort when it is unnecessary. 
Several methods have been proposed to solve this problem \cite{peng1996review}. All of them try to put some limit on the integral component.
We put two conditions on our integral component for direction $i$.
First, error rate $e(r)$ in the direction $i$ should be larger than $0$.
Second, we put a saturation point for how large the integral component can be.
These two conditions can help to reduce the integral windup effects.

\subsection{Other methods}

\par To analyze the performance of our PID-inspired method, we compare it to four different threshold modification methods: one static (no threshold modification) and three dynamic threshold models.

\par All four initial thresholds for each agent are drawn from a uniform distribution. We chose this initial distribution because it was shown in previous studies \cite{wu2020effects} that swarms with initial thresholds drawn from the uniform distribution perform better than those with gaussian, possion, and constant distributions.

\par The first method we compare our results to is TD0. With TD0, thresholds remain constant throughout the experiment. This method serves as a benchmark for other methods.

\par The second method, TD1,  is inspired by one of the most frequently used methods in response threshold studies: the learning/forgetting method \cite{theraulaz1998response}. The idea behind this method is that as agents gain experience in a certain task, they should become more likely to perform it in the future and less likely to perform the other tasks. In time step $t$, if agent $i$ performs task $k$, $\theta_{t,i,k}$ can be calculated by :
\begin{equation}
\theta_{t,i,k}=\theta_{t-1,i,k} - \epsilon \quad  k \in \lbrace N, E, S, W\rbrace
\end{equation}
All of the other task thresholds, $\theta_{t,i,j}$, $j\neq k$, for agent $i$, at time step $t$, can be written as:
\begin{equation}
\theta_{t,i,j}=\theta_{t-1,i,j} + \psi \quad
j \in \lbrace N, E, S, W\rbrace and j\neq k
\end{equation}

\par Using this method, the agent can specialize on specific tasks that are in line with the current task demands, but once the thresholds converge, agents have a hard time re-specializing to new sets of task demands \cite{kazakova2018specialization,kazakova2020respecializing}. Agents tend to get stuck with very high or very low thresholds, which we call sink-states.

\par The third method, TD2, is similar to TD1 since it utilizes the same learning/forgetting method previously explained. The only difference is that it tries to avoid the sink state problem of the TD1 method by introducing heterogeneous threshold ranges. 
Previous studies have suggested that using heterogeneous threshold ranges can help us avoid sink states \cite{wu2020dynamic}.
In TD1, all agents have the same threshold range, but in TD2, the minimum threshold value is generated randomly
from $[0, 0.5]$, and the maximum value is generated randomly from $[0.5,1]$ \cite{wu2020dynamic}. 

\par The fourth method, TD3, is very similar to TD2, with the difference being that threshold ranges are randomly drawn from [0,1]. The larger value is assigned to the maximum value of the range, and the smaller value is assigned to the minimum value of the range.

\par We want to test our PID-inspired method against the methods above to see how it performs and if it can avoid sinks-states and re-specialize effectively.

\section{Results and Discussion}

We test our method, as well as TD0, TD1, TD2, and TD3, on the straight, zigzag, sharp, s-curve, and random paths to examine its effectiveness on various task demand changes, and on multiple swarm sizes (50, 100, and 500 agents) to examine its scalability. Our PID-inspired method demonstrates superior performance compared to others across all paths and for all swarm sizes.
\par
\begin{figure*}[t]
  \centering
  
  \subfigure(a){\includegraphics[width=0.30\linewidth]{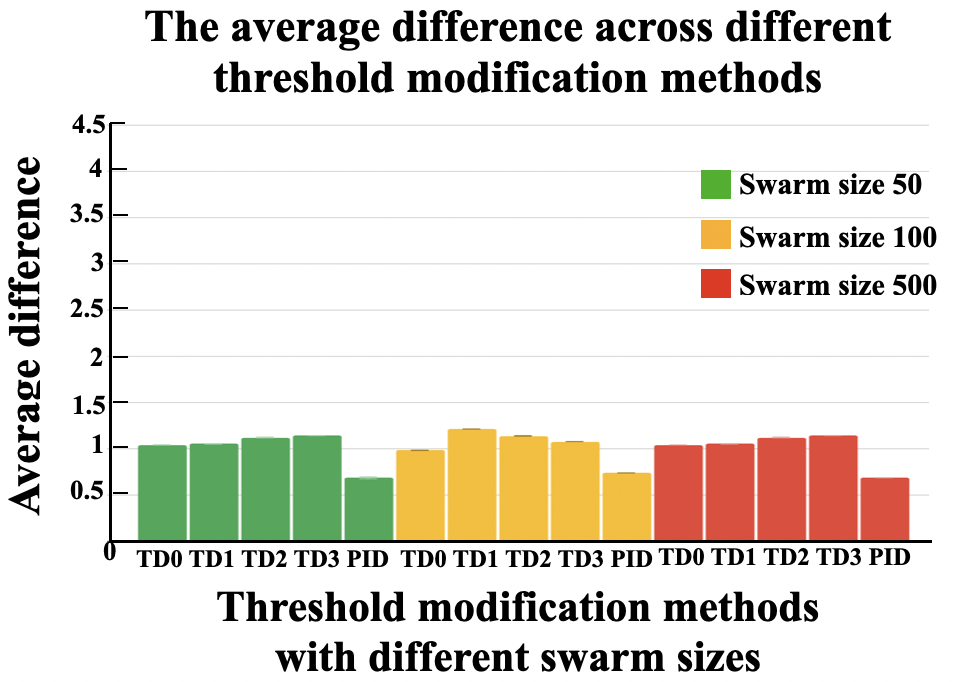}}
  \subfigure(b){\includegraphics[width=0.30\linewidth]{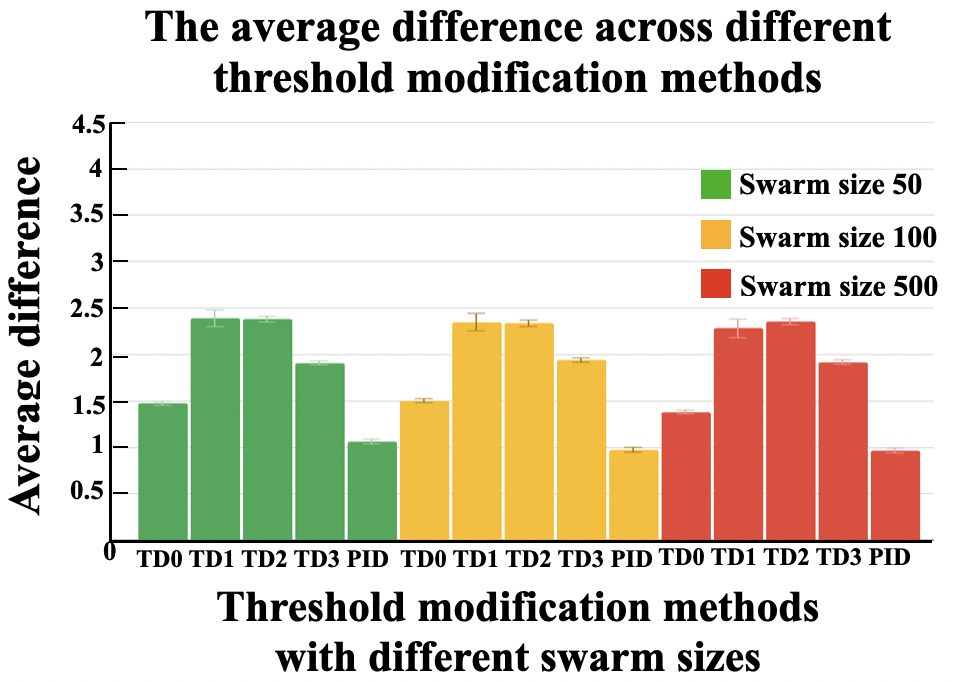}} 
  \subfigure(c){\includegraphics[width=0.30\linewidth]{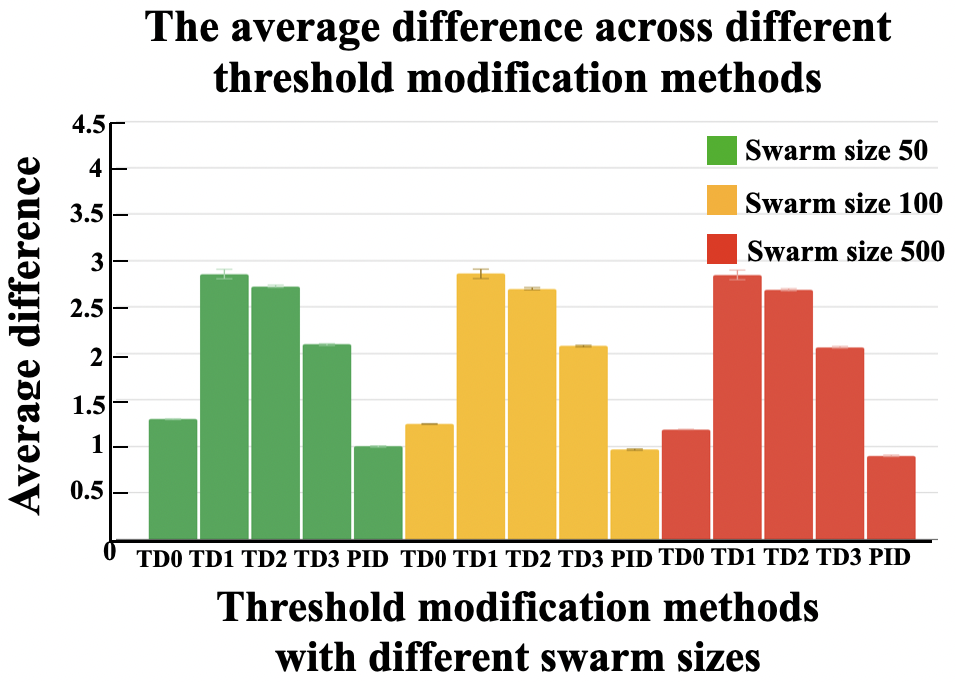}}\\
 \subfigure(d){\includegraphics[width=0.30\linewidth]{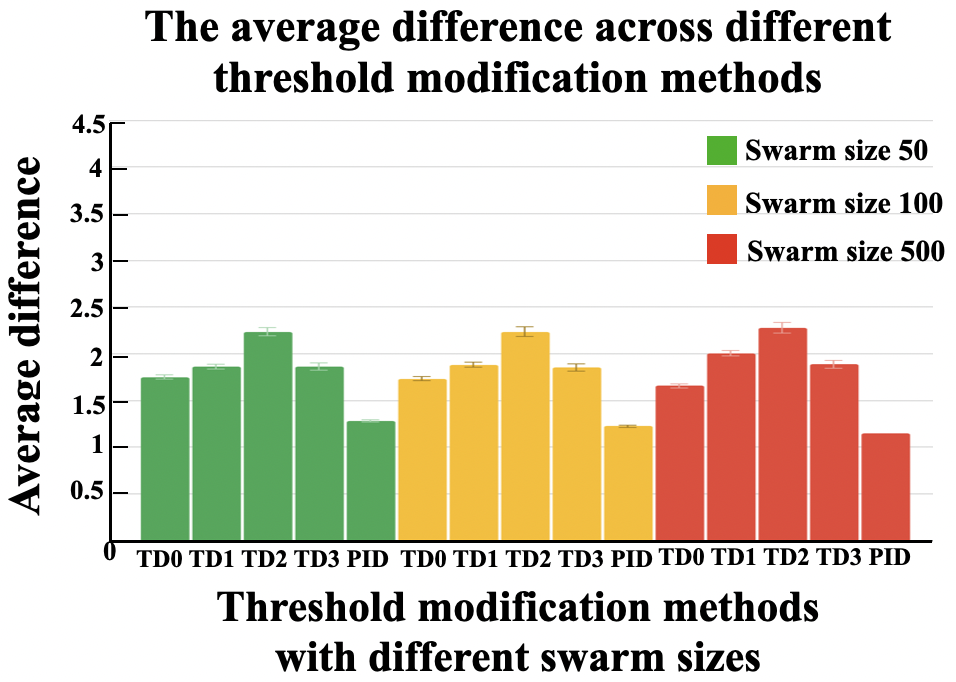}}
\subfigure(e){\includegraphics[width=0.30\linewidth]{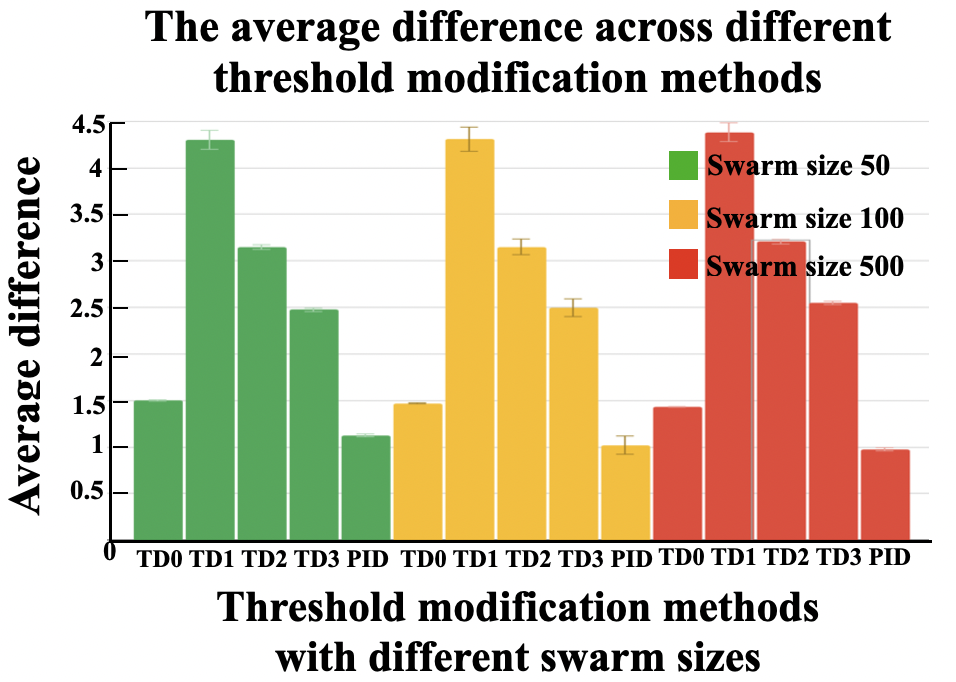}}
  \caption{Average distance between the target and the tracker averaged over 100 runs. Green indicates swarm size 50, yellow 100 and red 500 in a (a) Straight path (b) S-curve path (c) Random path (d) Zigzag path (e) Sharp path}
 
\label{fig:fig3}
\end{figure*}

 Figure~\ref{fig:fig3} compares the error rates of different threshold modification methods with different swarm sizes and task demand paths along with error bars representing one standard deviation. The smaller the average difference better the performance. The results are averaged out over 100 runs.
  Graph(a) depicts the error rate in a straight path, graph(b) in a s-curve path, graph(c) in a random path, graph(d) in a sharp path, and graph(c) in a zigzag path. The $y$-axis of each of the graphs represents the average error rate of the system throughout a run, averaged over 100 separate runs. As discussed before, in our setup, error rate is the difference between the target and the tracker. The $x$-axis of each graph represents different threshold modification methods (TD0, TD1, TD2, TD3, and PID-inspired method) with different swarm sizes. The color green represents swarm size 50, yellow represents swarm size 100, and red represents swarm size 500. Figure~\ref{fig:fig3} demonstrates that the PID-inspired method consistently outperforms other methods in various paths and with different swarm sizes. This indicates that our method is a non-risky choice for unknown paths and is scalable, a crucial factor for swarm system algorithms.

\subsection{Specialization analysis}
Specialization is beneficial to any multi-agent system \cite{whiten2022emergence}; we all use specialization, from insect societies \cite{beshers2001models} to human societies \cite{yang1998specialization}. 
Specialization is not an easy term to define. Even in response
threshold studies, there is no agreed-upon rule for when to call an agent specialized on a specific task. 
In our system, for a more thorough comparison of the methods, we need to examine how the thresholds of the system change and how they adapt to varying task demands.
Therefore, we call an agent a specialist in a particular task if that task's corresponding threshold is the lowest among the four. The speciality of agent $i$ , $S(i)$, in time step step $t$, given north threshold, $T_{i,North}$, east threshold, $T_{i,East}$ , south threshold, $T_{i,South}$ and west threshold, $T_{i,West}$ can be written as :
\begin{equation}
 S_t(i)=\arg\min\lbrace T_{i,North}, T_{i,East}, T_{i,South}, T_{i,West} \rbrace
\end{equation}

When changes in task demand are significant, a high variance for the number of agents specialized in a specific task indicates the system's responsiveness. When task demands are constant, low variance indicates the system's stability.

For a more thorough comparison of the methods, we examine examples of individual runs and how the agents' specialty changes.
Figures ~\ref{fig:4} to ~\ref{fig:8} represent how agents' specialization, $S_t(i)$, change over time in different paths. The $y$-axis in all figures represents the time step in the experiment. The $x$-axis represents $Agent_{0}$ through $Agent_{99}$. Each color in the graphs represents the agent's specialty at that time step. Blue shows north specialty; yellow shows east; red shows south; and green shows west.

\subsubsection{No change (Straight path)}

\begin{figure}[t]
  \centering
  \includegraphics[width=\linewidth]{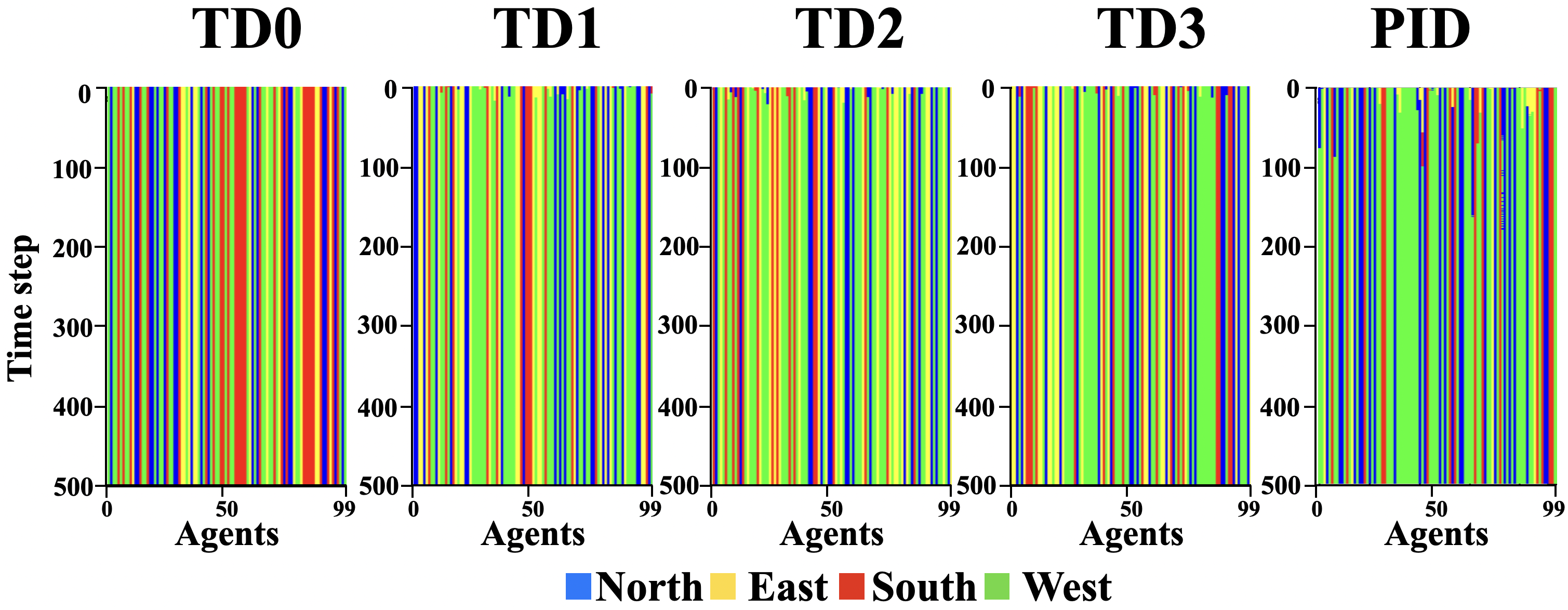}
  \caption{
Straight path specialization comparison of different threshold adaptation methods}
 \label{fig:4}
\end{figure}

We start with the simplest path. The straight path has a constant west task demand. 
As we have seen in Figure~\ref{fig:fig3}, the PID-inspired method outperforms TD0, TD1, TD2, and TD3 across all swarm sizes. 
In Figure~\ref{fig:4}, we can see that since the initial threshold for all agents is set randomly, the agent's specialty, $S(i)$, has a random pattern in the first few time steps. In the TD0 method, the thresholds are fixed, so we see no change in specialty across time, which is expected.

In TD1, TD2, TD3, and PID-inspired methods, we can see that some agents change their specialty,$S(i)$ from other tasks to west in the initial time steps; however, we witness that there are more agents with a west specialty in the PID-inspired method than others. The variance of the number of agents specialized in the west, in this particular run with the PID-inspired method, is $13.00$, which is higher than $0$, $3.06$, $4.33$, and $1.97$, which are the variance of the west specialists for TD0, TD1, TD2 and TD3, respectively. In the initial timesteps, agents become specialized in the west direction, and the higher variance number indicates that specialization on the west has occurred more with the PID-inspired method. If we only consider the last 100 timesteps, all variances go to zero, which indicates that agents settle and stablize after some time. This stabilization is expected since task demands are constant for the straight path.

\subsubsection{Constant change (s-curve path and random path)}

 Second, we analyze paths with gradual, constant changes in task demands, the s-curve, and random paths. The PID-inspired method outperforms the others in terms of the average distance between the target and the tracker.

\begin{figure}[t]
  \centering
  \includegraphics[width=\linewidth]{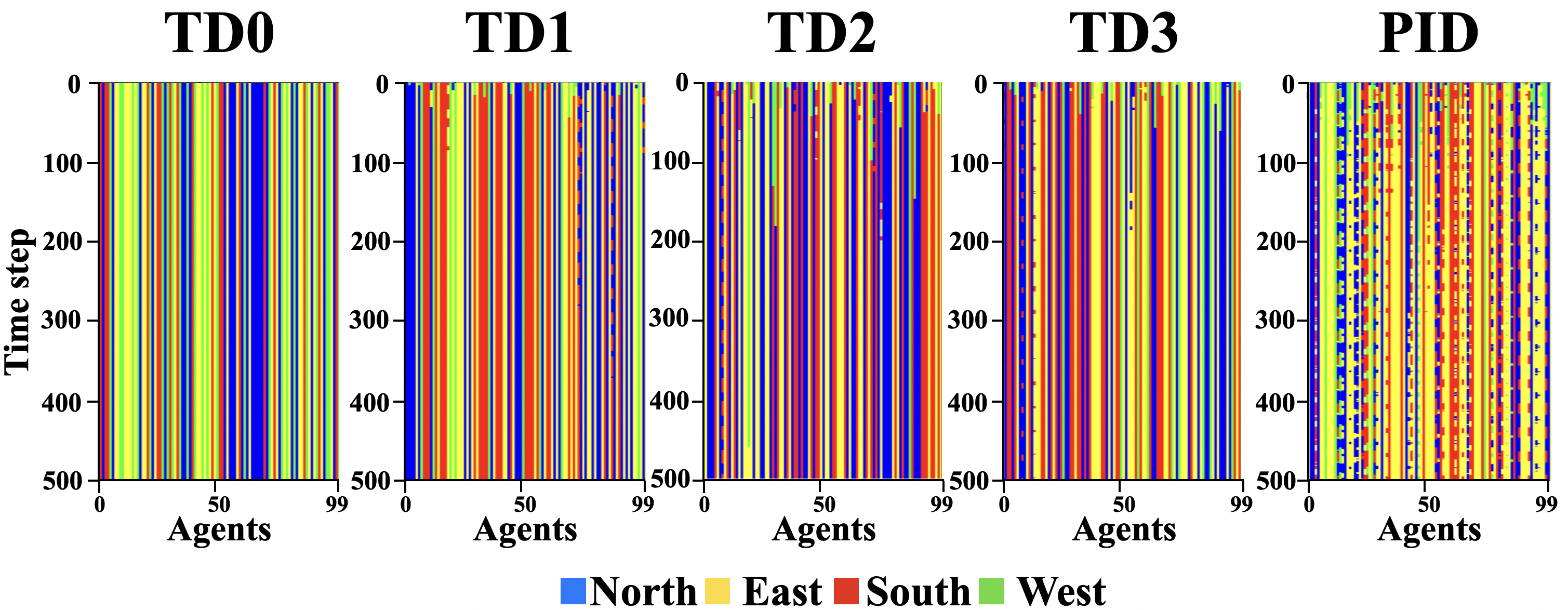}
  \caption{S-curve specialization comparison of different threshold adaptation methods}
  \label{fig:5}
  \Description{Specialization comparison of different threshold adaptation methods in a s-curve path}
\end{figure}

\begin{figure}[t]
  \centering
  \includegraphics[width=\linewidth]{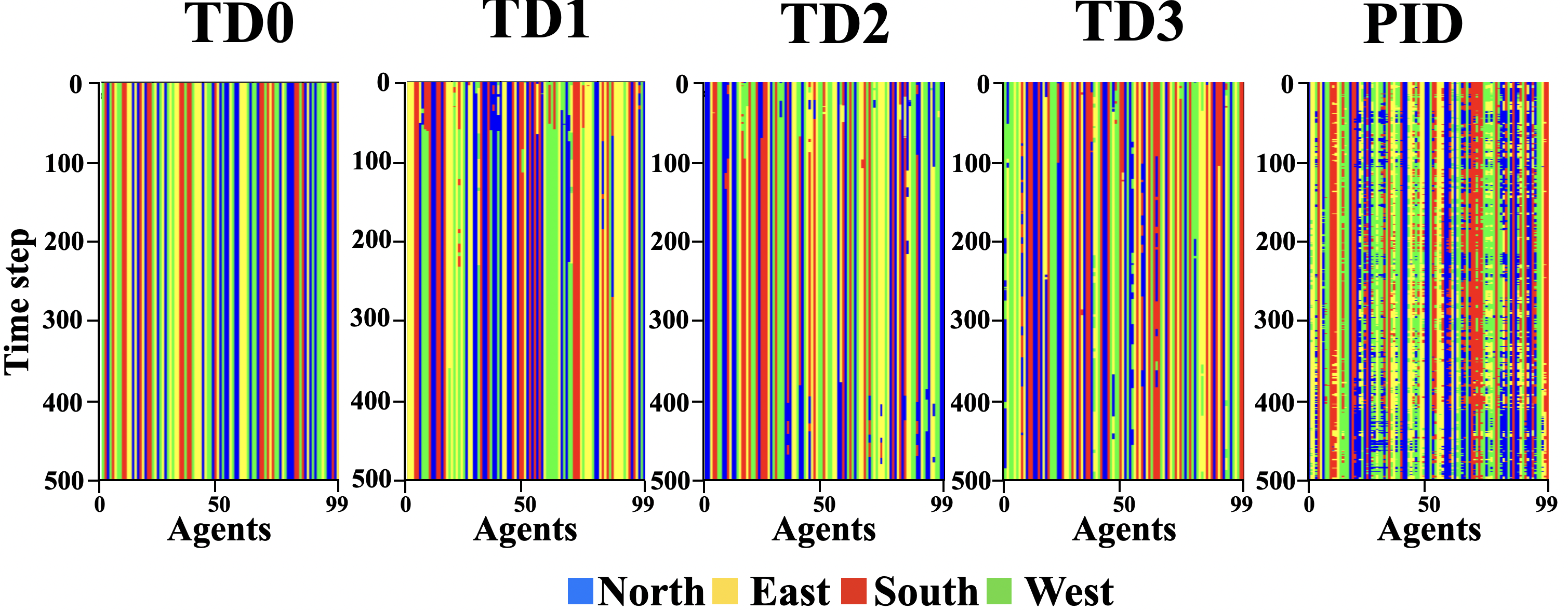}
  \caption{Random path specialization comparison of different threshold adaptation methods}
  \label{fig:6}
  \Description{Comparison of different threshold adaptation methods in a random path}
\end{figure}

Figure~\ref{fig:5} depicts the specialty graph for the s-curve path.
TD0 has a constant specialty through time. We can see that other methods specialize and re-specialize in earlier time steps; however, TD1, TD2, and TD3's ability to re-specialize diminishes as the experiment progresses. They become stuck in their previous specialty and fail to adapt. The PID-inspired method, however, can adapt to changing task demands. We can see a periodic pattern in the specialty change of its corresponding agents, which is in line with the s-curve's periodic pattern.
Figure~\ref{fig:6} shows the specialty graph for the random path. Again we can see that TD1, TD2, and TD3 have a harder time respecializing than the PID-inspired method.
\par Since in the s-curve and random paths, we have constant task demand changes; specialties never settle with the PID-inspired method. The PID-inspired method has higher variance in all task demands; for example, the variance for specialists in the north for the PID-inspired method in the s-curve path is 34.69, and for others, it is 0, 1.35, 2.87, and 0.69. For the random path, for the PID-inspired method, it is 67.52, and for others, it is 0, 2.40, 9.512, and 5.06, respectively. If we only consider the last 100 timesteps, variances remain similar to considering all 500 timesteps. This indicates that agents don't get stuck in sink states and stay responsive throughout the experiment.

\subsubsection{Abrupt change (Zigzag path and sharp path)}

\begin{figure}[t]
  \centering
  \includegraphics[width=\linewidth]{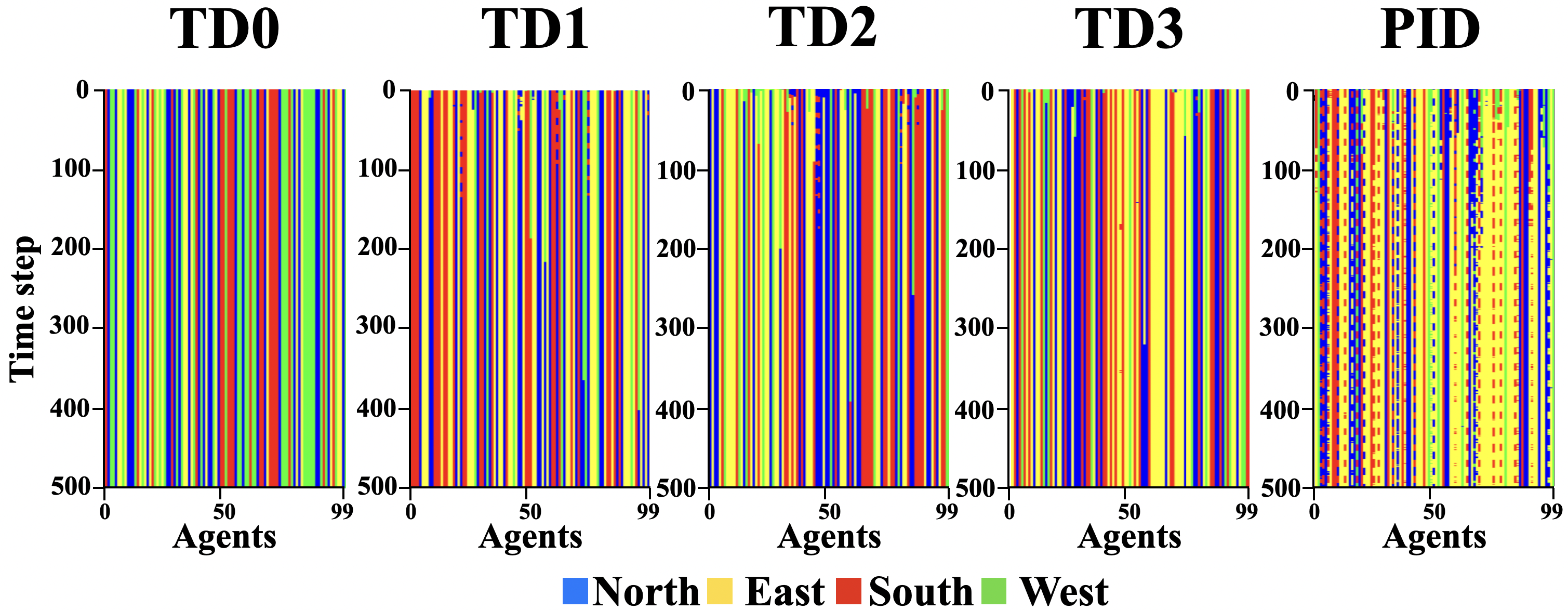}
  \caption{Zigzag specialization comparison of different threshold adaptation methods}
  \label{fig:7}
  \Description{Comparison of different threshold adaptation methods in a zigzag path}
\end{figure}

 Finally, we analyze paths that are constant for a while and then abruptly
change. The zigzag pattern has a constant east task demand, but north and south demand change periodically at peak points. In Figure~\ref{fig:7}, we can see again that TD1, TD2, and TD3 specialize and re-specialize in the early stages, but as time goes by, they become less responsive but PID-inspired method can continue to adapt.

\begin{figure}[t]
  \centering
  \includegraphics[width=\linewidth]{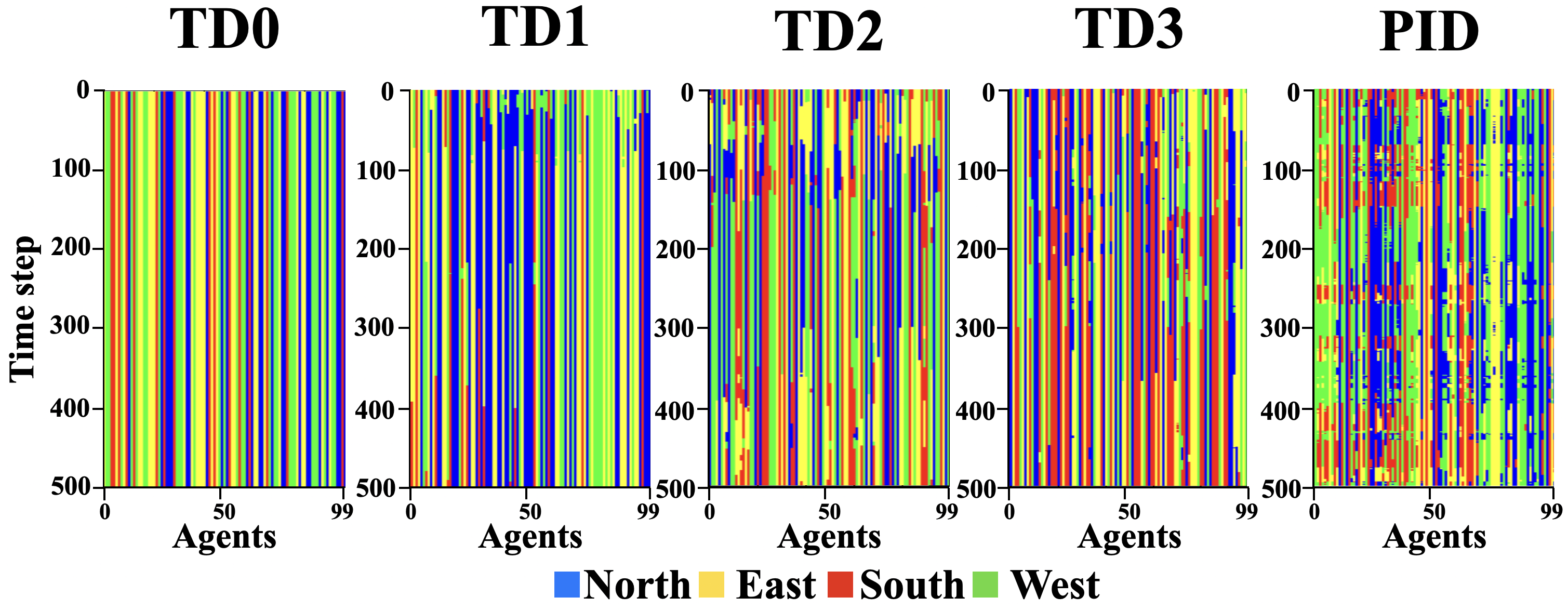}
  \caption{Sharp path specialization comparison of different threshold adaptation methods}
  \label{fig:8}
  \Description{Comparison of different threshold adaptation methods in a sharp path}
\end{figure}

For the sharp path, the task demand stays constant and suddenly takes a random turn. In Figure~\ref{fig:8}, we can see again that even with a non-periodic path,  the PID-inspired method has a better ability for respecializing than other methods.
In the zigzag and sharp path, task demands stay constant for a while and then abruptly change. If we consider the 500 timesteps, the variance of specialists for all tasks is higher in the PID-inspired method. For example variance of the number of specialists in the south direction for the PID-inspired method in the zigzag path is 36.79, and the others are 0, 1.47, 2.53, and 0.28. If we only consider the last 100 timesteps, PID-inspired method variances reduce significantly, for example in the sharp path variance of the south direction speciality in 500 time steps is $43.61$, and in 100 timesteps it is $8.02$. That is because task demands in the sharp and zigzag paths remain constant for a while and then abruptly change. We have a low variance for periods when task demand is constant, and agents' specialties do not change as much, indicating systems stability under constant task demand.

\begin{figure}[t]
  \centering
   \subfigure(a){\includegraphics[width=8cm]{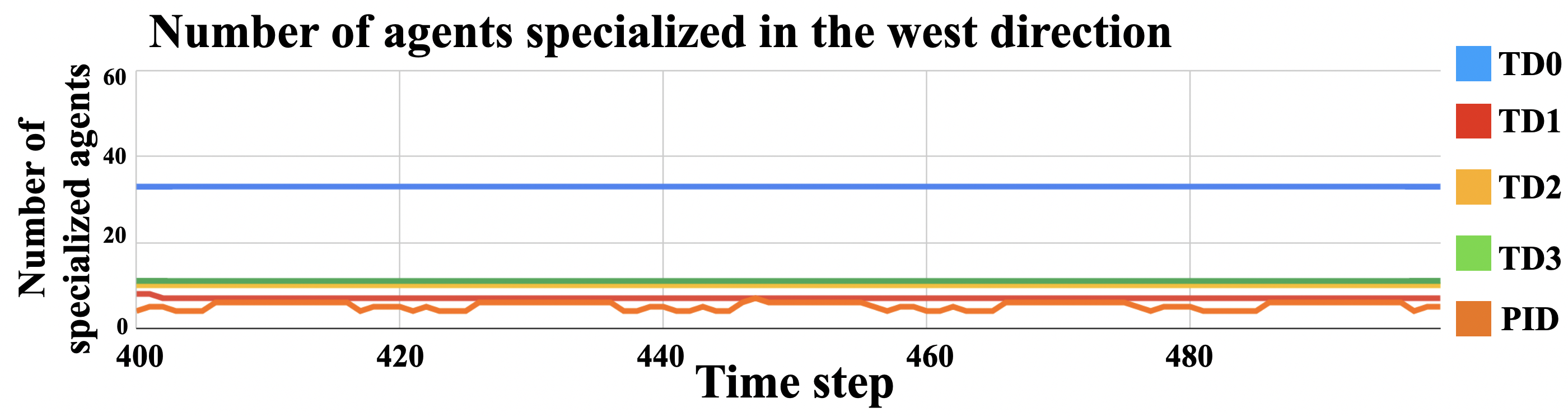}}\\
  \subfigure(b){\includegraphics[width=8cm]{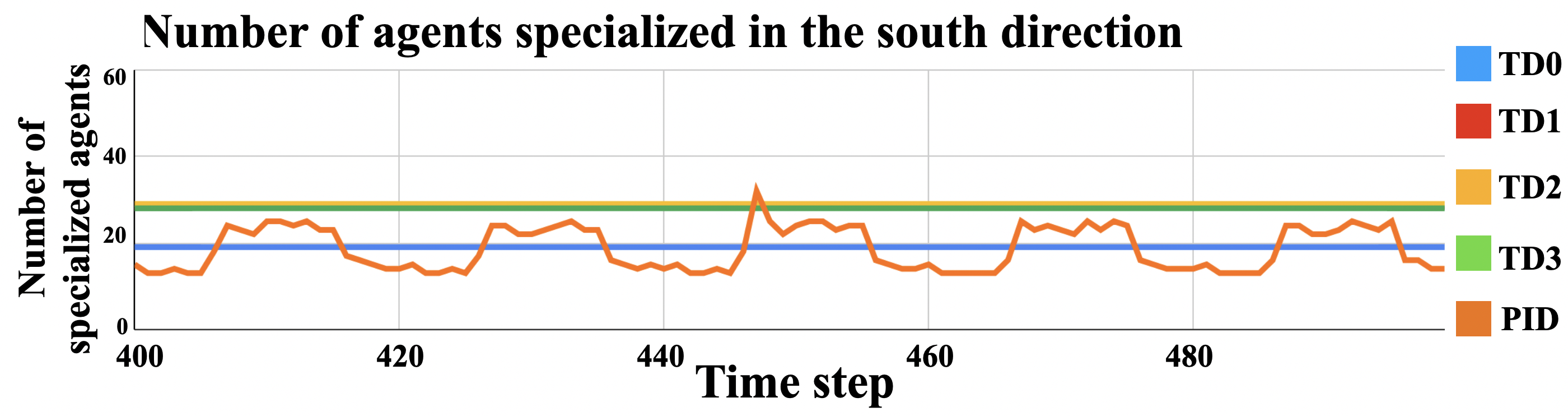}} \\
  \subfigure(c){\includegraphics[width=8cm]{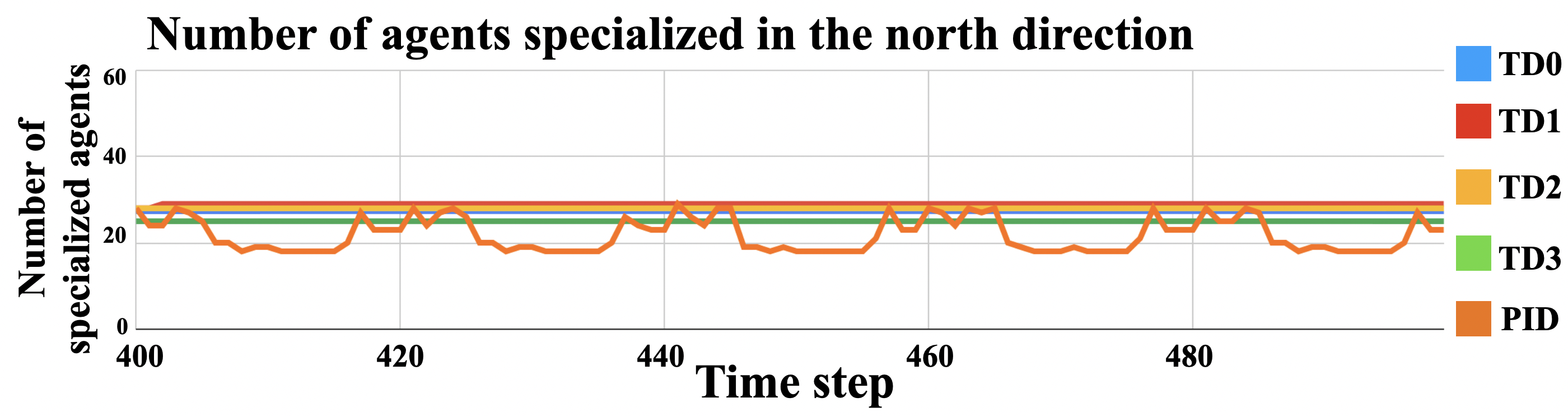}}\\
  \subfigure(d){\includegraphics[width=8cm]{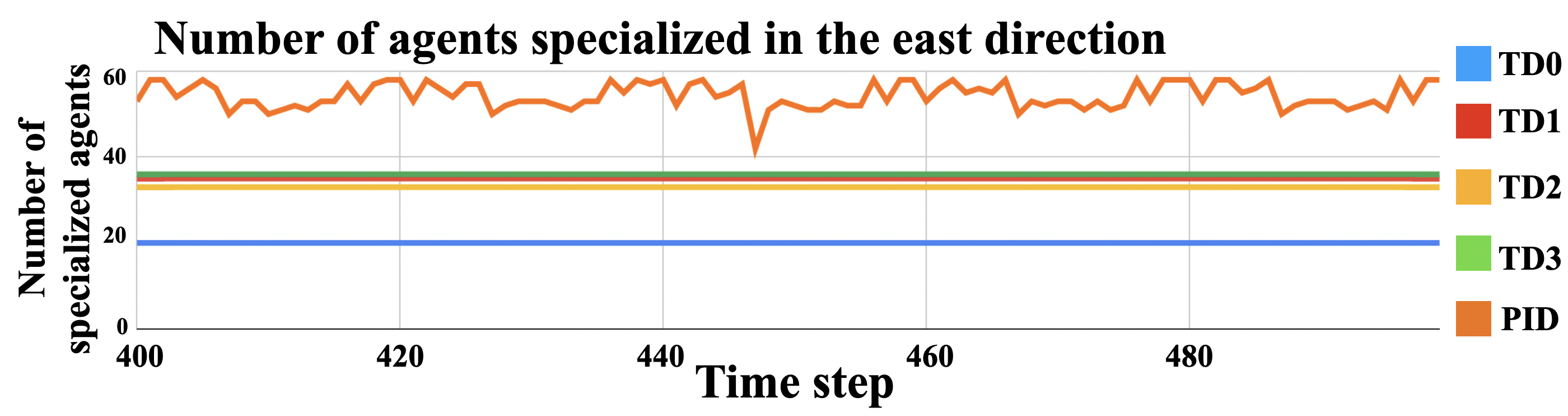}}
  \caption{Number of specialized agents with different threshold modification methods from timestep 400 to 500 in task: (a) West direction (b) South direction (c) North direction (d) East direction }
 
\label{mm}
\end{figure}

To illustrate the effectiveness of specializations and respecialization, we discuss an example from the zigzag path. We choose the zigzag path because it has explicit points for abrupt changes in task demands.
Figure~\ref{mm} shows the number of specialized agents on each task with different threshold modification methods.
A small time interval [400,500] is chosen for a clearer depiction. The $x$-axis shows time, and the $y$-axis indicates the number of agents specialized in tasks (a) west, (b) south, (c) north, and (d) east. Blue represents TD0; red, TD1; yellow, TD2; green, TD3; and orange, PID. 
In the zigzag path, task demands for north and south alternate periodically. The same pattern can be seen in Figure~\ref{mm}c and Figure~\ref{mm}b.
The zigzag path takes the northeast direction first, which explains why there is a bump at the beginning of Figure~\ref{mm}c and Figure~\ref{mm}e. Agents specialize in the north and east direction initially, but after the south task demand increases, some of the agents with previous north and east specialty change to south specialty. This is why we see similar highs and lows for the north and east. The number of west specialists has very small changes since there is no task demands in that direction. The small changes are due to the system trying to correct small over-shootings at peak points when task demands change abruptly. These over-shootings can be reduced by tuning the parameters on zigzag path itself instead of the straight path or using more precise tuning methods.
These figures show that the specializations of the agents are in line with the task demands and that PID is more responsive and is able to specialize and re-specialize.
 
\section{Conclusion}

In this study, we test the effectiveness of a novel PID-inspired threshold modification method on a decentralized, non-communicating, threshold-based swarm system. We test our method on a collective tracking problem in a 2D vector space. We consider a target and a tracker that can move a fixed distance at each time step and the tracker is collectively controlled by the swarm of agents.
Prior threshold modification methods can effectively specialize initially, but the agents have a hard time re-specializing when demands change and fail to adapt. 
Our method is based on the PID control loop factors, which are the error rate, the derivative of the error rate, and the integral of the error rate. The error rate in our system is defined as the distance between the target and the tracker. Each agent has a separate controller that modifies the thresholds of that agent.
Our results indicate that our method outperforms other discussed methods across all paths regardless of the swarm size. Under constant task demand, the agents reach stable specialization, and with changing task demands, agents can specialize and re-specialize without becoming stuck. Our method is immune to sink states and is, therefore, able to re-specialize to changing task demands throughout a simulation.

\section{Acknowledgments}
This work was supported by the National Science Foundation under Grant
No. IIS1816777.

%%
%% The next two lines define the bibliography style to be used, and
%% the bibliography file.
\bibliographystyle{ACM-Reference-Format}
\bibliography{sample-base}

%%
%% If your work has an appendix, this is the place to put it.
\appendix

\end{document}